\documentclass[twocolumn,showpacs,showkeys,preprintnumbers,amsmath,amssymb]{revtex4}

\usepackage{graphicx}

\usepackage{dcolumn}

\usepackage{bm}

\begin{document}


\title{Scalable Percolation Search in Power Law Networks}


\author{Nima Sarshar}
\author{P. Oscar Boykin}

\author{Vwani Roychowdhury}

\email{{nima,boykin,vwani}@ee.ucla.edu}

\affiliation{Department of Electrical Engineering, University of California,
Los Angeles}%



\begin{abstract}
We introduce a scalable searching algorithm for finding nodes and
contents in random networks with  Power-Law (PL) and heavy-tailed
degree distributions. The network is searched using a
probabilistic broadcast algorithm, where  a query message is
relayed on each edge with probability just above the bond
percolation threshold of the network. We show that if each node
caches its directory via a short random walk, then the total
number of {\em accessible contents exhibits a first-order  phase
transition}, ensuring very high hit rates just above the
percolation threshold. In any random PL network of size, $N$, and
exponent, $2 \leq \tau < 3$,  the total traffic per query scales
sub-linearly, while the search time scales as $O(\log N)$.   In a
PL network with exponent, $\tau \approx 2$, {\em any content or
node} can be located in the network with {\em probability
approaching one} in time $O(\log N)$, while generating traffic
that scales as $O(\log^2 N)$, if the maximum degree, $k_{max}$, is
unconstrained, and as $O(N^{\frac{1}{2}+\epsilon})$ (for any
$\epsilon>0$) if $ k_{max}=O(\sqrt{N})$. Extensive large-scale
simulations show these scaling laws to be precise. We discuss how
this percolation search algorithm can be directly adapted to solve
the well-known scaling problem in  unstructured Peer-to-Peer (P2P)
networks. Simulations of the protocol on sample large-scale
subnetworks of existing P2P services show that overall traffic can
be reduced by almost two-orders of magnitude, without any
significant loss in search performance.

\end{abstract}


\maketitle

\section{Introduction and Motivation}\label{I}

Scale-free networks with heavy-tailed and Power-Law (PL) degree
distributions have been observed in several different fields and
scenarios (see, e.g.,  \cite{albe02} and references therein).   In a PL
degree distribution, the probability that a randomly chosen node
has degree $k$ is given by $P(k) \sim k^{-\tau}$, where $\tau>0$
is referred to as the exponent of the distribution.  For $2\le
\tau \le 3$  a network with $N$ nodes has constant or at most
$O(\log N)$ average degree, but the variance of the degree
distribution is unbounded. It is in this regime of $\tau$ that the
PL networks display many of the advantageous properties, such as
small diameter \cite{Newman}, tolerance to random node deletions
\cite{CNSW00}, and a natural hierarchy, where there are
sufficiently many nodes of high degree.

The searching problem in  random power-law networks can be stated as
follows\cite{hub}: Starting from a randomly selected node, {\em the source}, find another randomly selected node, {\em the destination}, through only {\em local
communications}. Equivalently, this can be cast into a messaging
problem, where it is desirable to transfer a message from an
arbitrary node to another randomly chosen node through local
(i.e., first neighbor) communications. Since a searcher has no
idea about the location of the destination node in the network
(unless, each node somehow  has path information for all other
nodes cached in it), the problem is indeed that of transferring a
message from a node to \emph{all} other nodes in the network.

Another equivalent version of this problem appears in {\em unstructured P2P
networks}, such as Gnutella\cite{rfc-gnut}, Limewire\cite{limewire},
Kazaa\cite{kazaa}, Morpheus\cite{morpheus}, and Imesh \cite{imesh}, where the
data objects {\em do not have global unique ids}, and {\em queries are done via
a set of key words}. The reasons that search in PL networks is important for
such unstructured P2P networks,  include: (i) A number of recent studies have
shown that the structure of these existing networks has complex network
characteristics \cite{Sar02,Ripeanu02}, including approximate power law degree
distributions. Thus PL networks, or at least networks with heavy-tailed degree
distributions, seem to naturally emerge in the existing services. (ii)
Systematic P2P protocols that will lead to the emergence of PL networks with
tunable exponents, even when nodes are deleted randomly, have been proposed
recently \cite{us}. This makes it possible to systematically design robust and random P2P
networks that admit PL degree distributions, and that can exploit several properties of PL graphs that are extremely useful for networking services, e.g., low diameter, which allows fast searches, a randomized hierarchy, which
allows optimal usage of heterogeneous computing and networking resources without the intervention of a global manager, and
extreme tolerance to random deletions of nodes, which provides robustness.

In a straightforward parallel search approach in P2P networks,
each query is given a unique id, and then each node on receiving
the query message sends it out to all of its neighbors, unless the
node has already processed the query, which the node can identify
by checking the id's of the queries it has already processed. This
leads to $O(N)$ total queries in the network for every single
query, and results in significant scaling problems. For example,
Ripeanu et.al.\cite{Ripeanu02} estimated that in December of 2000
Gnutella traffic accounted for $1.7\%$ of Internet backbone
traffic.

 As reviewed later in Section \ref{p2p-sec}, a number of
ad hoc measures, ranging from forcing an ultra-peer structure on
the network, to  a random-walk based approach, where it is assumed
that a constant fraction of the nodes in the network caches each
content's location, have been proposed in the P2P community. But
none of these measures provides a provably scalable and
decentralized solution, where any content, even if it is located
in only one node, is guaranteed to be found. The only systematic
work on searches in random PL networks  reported so far
\cite{hub}, employs a {\em serial search technique} based on
random walks and caching of content-lists of every node on all its
neighbors (or on all its first and second neighbors), and is
reviewed in greater detail in Section \ref{prior-work}.

We present a parallel, but scalable,  search algorithm that
exploits the structure of PL networks judiciously, and provides
precise scaling laws that can be verified via extensive
large-scale simulations (Section \ref{III}).  The key steps in our
search algorithm are: (i) {\em Caching or Content Implantation:}
Each node executes a short random walk and caches its content list
or directory on the visited nodes. For example, for $\tau\approx
2$, this one-time-only random walk is of length $O(\log N)$, and
thus the average cache size per node is $O(\log N)$. (ii) {\em
Query Implantation}: When a node wants to make a query, it first
executes a short random walk and implants its query request on the
nodes visited. (iii) {\em Bond Percolation:} All the implanted
query requests are propagated independently through the network
{\em in parallel} using a probabilistic broadcast scheme. In this
scheme, a node on receiving a query message for the first time,
relays the message on each of its edges with a certain probability
$q$, which is vanishingly greater than the percolation threshold,
$q_c$, of the underlying PL network (see \cite{sbr-percolation}
for a review of bond percolation in PL graphs).

The  physics of how and why percolation search algorithm works
efficiently, can be described as follows. The bond percolation
step, executed just above the percolation threshold, guarantees
that a query message is received by all nodes in a giant connected
component of diameter $O(\log N)$ and consisting of high-degree
nodes.   The content and query implantation steps ensure that the
content list of every node is cached on at least one of the nodes
in this giant component with probability approaching one, and that
one of the nodes in the giant connected component receives a query
implantation with probability approaching one.  Thus with
$O(\langle k \rangle N q_c)$ traffic (which scales sublinearly for
PL graphs, as shown in Section \ref{III} and
\cite{sbr-percolation}), any content ({\em even if it is owned by
a single node in the network}) can be located with {\em
probability approaching one} in time $O(\log N)$.

An interesting outcome pertaining to the physics of networking is that the  accessible contents/nodes
{\em exhibit a first-order phase transition} as a function of the broadcast or
percolation probability $q$, showing a sharp rise as soon as $q$ exceeds the
percolation threshold $q_c$. In contrast to the accessible contents, the {\em
number of nodes and edges} in the giant connected component {\em exhibits only
a  second order phase transition}. One of the {\em primary appeals of the
percolation search algorithm} is that by combining serial random walks (i.e.,
content and query implantations) with bond percolation {\em it engineers a
second-order phase transition into a first-order}, allowing query-hits
approaching 100\%, even when $\displaystyle \lim_{N\rightarrow \infty}(q-q_c)
=0$.

While the proof that the percolation search algorithm leads to
scalable traffic and low latency is based on fairly involved
concepts, the {\em algorithm itself can be easily implemented} and
{\em directly adapted to solve the scaling problem plaguing
unstructured P2P networks}. In Section \ref{p2p-sec}, we discuss
such applications, and present simulation results which show that
even on sample large-scale subnetworks of existing P2P services,
the overall traffic can be reduced by almost two-orders of
magnitude, without any significant loss in search performance, by
a direct implementation of percolation search.  We also consider
{\em heterogeneous networks} in Section \ref{hetero-sec}, where
the degree distribution is a mixture of heavy-tailed and
light-tailed PL distributions. Such mixture distributions can
model networks, such as the popular P2P services, where nodes
belong to only few types, and each type has its own capability and
hence, its own degree distribution. We provide both simulation and
analytical studies of the improvements to be accrued from the
percolation search algorithms when implemented on random
heterogeneous networks (Section \ref{hetero-sec}).

\section{Prior Work and Comparison}\label{prior-work}
 The search algorithm by Adamic et. al.
\cite{hub} can be described as follows: To convey a message from
node $A$ to $B$, $A$ sends a message that goes on a random walk
through the network. When arriving at a new node, the message
requests it to scan all its neighbors for the destination node
$B$. If $B$ is not found among the neighbors of the current node, then the message
is sent to one of the neighbors of the current node picked randomly.

This algorithm exploits the skewed degree distribution of the
nodes in PL networks: The random walk naturally gravitates towards
nodes with higher degree, and therefore, by scanning the neighbors
of these high degree nodes, the random walker is expected to soon
be able to scan a large fraction of the network. One could also
{\em scan both the first and the second neighbors} of a node
visited through the random walk (rather than just scanning its
first neighbors), in order to find the destination node $B$.

Estimates for both search time and the number of messages created (i.e., traffic)  per query can be obtained as follows:
For a power-law random graph with exponent $\tau$, the expected
degree of a node arrived at via a random link is $z_{a} \propto
k_{max}^{3-\tau}\propto N^{\frac{3}{\tau}-1}$, assuming that
$k_{max}=N^{1/\tau}$. Also, the expected number of the second
neighbors of a node randomly arrived at by following a link is
around $z_{b}\propto N^{2(\frac{3}{\tau}-1)}$. Therefore, assuming that nodes are not scanned multiple times
during the random walk, the whole network is expected to be
scanned after around:
\begin{equation}\label{slaws}
 A_a\approx N/z_a=N^{2-\frac{3}{\tau}}
\end{equation}
hops if only the first neighbors are scanned, and
\begin{equation}\label{slaws2}
 A_b\approx N/z_b=N^{3-\frac{6}{\tau}}
 \end{equation}
  if the second neighbors are scanned as well. For $\tau=2$, and the case
where {\em both first and second neighbors} of a node are scanned, the
predicted scaling is poly-logarithmic in the size of the network.

While this technique is an {\em important first step towards exploiting the
hierarchical structure of PL networks} and provides a sublinear scaling of
traffic, there are several drawbacks that need to be addressed:
\begin{itemize}
\item[(i)]The actual performance of the algorithm {\em is far
worse than the theoretically predicted scaling laws}. The primary
reason for this discrepancy is that the estimates in
Eqs.~(\ref{slaws})~and~(\ref{slaws2}) are based on the assumption
that the nodes scanned during a walk are unique, i.e., no node is
scanned more than once. As pointed out by the authors in
\cite{hub},  while this is a good approximation at the start of
the walk, it quickly becomes invalid when a good fraction of the
nodes have been scanned. Extensive simulations in \cite{hub} show
that actual scaling is significantly worse than the predicted
values: For example, for $\tau=2.1$, Eq.~(\ref{slaws}) predicts a
scaling of $N^{0.14}$, but the actual scaling observed is  more
than a power of $5$ worse (i.e., $N^{0.79}$). The same is true for
(\ref{slaws2}), where a scaling of $N^{0.1}$ is predicted for
$\tau=2.1$ while $N^{0.71}$ is observed. \item[(ii)]{\em The
random-walk search is serial} in operation, and even assuming that
the predicted scalings are accurate, the search time for finding
any node or its content in the network is polynomially long in
$N$. As an example, for $\tau=2.3$, a value observed in early
Gnutella networks, the predicted search time scalings  are:
$A_a=N^{0.66}$ or $A_b=N^{0.39}$.

However, as mentioned before, these scalings are going to be
significantly worse and we know that they will be at least larger
than $N^{0.71}$.
\item[(iii)] In order to obtain the best traffic
scalings, one needs to scale cache (storage) size {\em per node}
polynomially; e.g., for $\tau\approx 2$, the cache size per node
should increase as $O(\sqrt{N})$. Recall that the search strategy
requires every node to answer if the node/content satisfying the
query message is in any of its first neighbors or in any of its
first and second neighbors. This scanning can be performed in two
ways: (1) {\em Without caching:} For each query message, the node
queries all its first (or first and second) neighbors. {\em This
strategy is then at least as bad as flooding}, since for each
independent search, all the links have to be queried at least once
which results in a traffic per search of at least $O(N)$. (2) {\em
With Caching:}   Each node caches its content-list on all its
neighbors, or on all of its first and second neighbors, as
required by the protocol. Through the random walk, the walker can
scan the contents of the neighbors (or both first and second
neighbors) by observing the content lists in the current node
without having to query the neighbors. The total cache size
required per node in the case of the first-neighbor-only caching
scheme is exactly the average degree of nodes (i.e., $O(\log N)$
for $\tau=2$), and $N^{3/\tau-1}$ (i.e., $O(\sqrt{N})$ for
$\tau=2$) when scanning of both the first and second neighbors are
required. Thus the least traffic and equivalently, shortest search
times, are obtained at the expense of an increased cache size
requirements per node.
\end{itemize}

As noted in the introduction and elaborated in the later sections, we build on
the basic ideas in \cite{hub}, and exploit the hierarchical structure of PL
networks more efficiently to successfully resolve many of the above-mentioned
issues. In particular, our results have the following distinctive features: (1)
The actual performance of the algorithm matches the theoretical predictions.
(2) The algorithm takes $O(\log N)$ time and is parallel in nature. (3) The
average cache size increases with the exponent $\tau$, and is minimum for
$\tau=2$, when the traffic scaling is the most favorable.
For example, for a random PL network with exponent, $\tau =2$, and maximum
degree $k_{max}$, we show that any content in the network can be found with
probability approaching one in time $O(\log N)$, while generating only
$O(N\times \frac{2\log k_{max}}{k_{max}})$ traffic per query. Moreover, the
content and query implantation random walks are $O(\log N)$ in size, {\em leading to the average cache size of $O(\log N)$}. Thus, if
$k_{max} = cN$ (as is the case for a randomly generated PL network with no a
priori upper bound on $k_{max}$) then the overall traffic scales
as $O(\log^2 N)$ per query, and if $k_{max} = \sqrt{N}$ (as is the usual
practice in the literature) then the overall traffic scales as $O(\sqrt{N}
\log^2N)=O(N^{\frac{1}{2}+\epsilon})$ (for any $\epsilon >0$) per query.

\section{The Percolation Search Algorithm and Its Scaling Properties}\label{III}
The percolation search algorithm can be described as
follows: \\
(i) {\em Content List Implantation:} Each node in a network of
size $N$ duplicates its content list (or directory) through a
random walk of size $L(N,\tau)$ starting from itself. The exact
form of $L(N,\tau)$ depends on the topology of the network (i.e.,
$\tau$ for PL networks), and is in general a sub-linear function
of $N$. Thus the total amount of directory storage space required
in the network is $N L(N,\tau)$, and the average cache size is
$L(N,\tau)$. Note that, borrowing a terminology from the Gnutella
protocol, {\em the length of these implantation random walks will
be also referred to as the TTL (Time To Live)}.  \\
(ii) {\em Query Implantation:} To start a query, a query request
is \emph{implanted} through a
random walk of size $L(N,\tau)$ starting from the requester. \\
(iii) {\em Bond Percolation:} When the search begins, each node
with a query implantation starts a {\em probabilistic broadcast
search}, where it sends a query to each of its neighbors with
probability $q$, with $q=q_c/\gamma$ where $q_c$ is the
percolation threshold\cite{sbr-percolation}.

We next derive scaling and performance measures of the above
algorithm. Our derivations will follow the following steps:

\begin{itemize}

\item First we define high degree nodes and compute the number of high
degree nodes in a given network.

\item Second, we show that after the probabilistic broadcast step
(i.e., after performing a bond percolation in the query routing
step), a query is received by all members of connected component
to which an implant of that query belongs.  We also see that the
diameter of all connected components is $O(\log N)$, and thus the
query propagates through it quickly.

\item Third, we show that  {\em a random walk of length
$L(N,\tau)$} starting from any node will {\em pass through a
highly connected node, with probability approaching one}. This
will ensure that (i) a pointer to  any content is owned by at least one highly connected node, and (ii) at least one
implant of any query is at one of the high degree nodes.

\item Finally, we examine the scaling of the maximum degree of the network
$k_{max}$ and give the scaling of query costs and cache sizes in terms of the
size of the entire network $N$.  We show that both cache size and query cost
scale sublinearly for all $2\le \tau < 3$, and indeed can be made to scale
$O(\log^2 N)$ with the proper choice of $\tau$ and $k_{max}$.
\end{itemize}

\subsection{High Degree Nodes}
\label{ssec:high_degree_nodes}
In this section we define the notion of a high degree node.  For any node with
degree $k$, we say it is a high degree node if $k \ge k_{max}/2$.
We assume that we deal with random power-law graphs which have a degree
distribution:
\begin{eqnarray*}
p_k&=& A k^{-\tau}, \\
\hbox{where }\, A^{-1} &=& \sum_{k=2}^{k_{max}} k^{-\tau}\approx \zeta (\tau) - 1,
\end{eqnarray*}
and $\zeta (\cdot)$ is the Riemann zeta function.
$A$ approaches the approximate value quickly as $k_{max}$ gets large, and thus
can be considered constant.
Thus the number of high degree nodes, $H$ is given by:
\begin{eqnarray*}
H &=& N \left( A \sum_{k=k_{max}/2}^{k_{max}} k^{-\tau} \right).
\end{eqnarray*}
Since for all decreasing, positive,$f(k)$ we have $\sum_{k=a}^b
f(k) > \int_a^{b+1}f(k)dk > \int_a^{b}f(k)dk$ and $\sum_{k=a}^b
f(k) < \int_{a-1}^b f(k)dk$, we can bound $H$ from above and
below:
\begin{eqnarray*}
H &>& \frac{A}{\tau -
1}\left(\frac{1}{(\frac{1}{2})^{\tau-1}}-1\right)\frac{N}{k_{max}^{\tau-1}}, \hbox{ and}\\
H &<& \frac{A}{\tau -
1}\left(\frac{1}{(\frac{1}{2})^{\tau-1}(1-\frac{1}{k_{max}/2})}-1\right)
\frac{N}{k_{max}^{\tau-1}} .
\end{eqnarray*}
For $k_{max}\rightarrow\infty$ we have that $\frac{1}{k_{max}/2}\rightarrow 0$
thus:
\begin{eqnarray*}
H \approx \frac{A}{\tau - 1}\left(2^{\tau -1} -
1\right)\frac{N}{k_{max}^{\tau-1}} .
\end{eqnarray*}
We have shown that $H = O(\frac{N}{k_{max}^{\tau-1}})$.
As we discuss in section \ref{ssec:on_maximum}, there are two choices for
scaling of $k_{max}$.  If we put no prior limit on $k_{max}$ it will scale
like $O(N^{1/(\tau - 1)})$.  As we will discuss, we may also consider
$k_{max}=O(N^{1/\tau})$.  We should note that the first scaling law gives
$H = O(1)$, or a constant number of high degree nodes as the
system scales.  The second gives $H = O(N^{1/\tau})$.  For all $\tau
\ge 2$, we have $H$ scaling sublinearly in $N$.

In the next sections we will show that without explicitly identifying or
arranging the high degree nodes in the network, we can still access them
and make use of their resources to make the network efficiently searchable.

\subsection{High Degree Nodes are in the Giant Component}
\label{ssec:high_degree_giant} In conventional percolation
studies, one is guaranteed that as long as $q-q_c=\epsilon >0$,
where $\epsilon$ is a constant independent of the size of the
network, then there will be a giant connected component in the
percolated graph. However, in our case, i.e., PL networks with
$2\leq \tau \leq 3$, $\lim_{N\rightarrow \infty} q_c =0$ (for
example, $q_{c}=\frac{\log(k_{max})}{k_{max}}$ for a PL network
with exponent $\tau=2$ \cite{Chung}), and since the traffic (i.e.,
the number of edges traversed) scales as $O(\langle k \rangle N
q)$, we cannot afford to have a constant $\epsilon>0$ such that
$q=\epsilon+q_c$: the traffic will then scale linearly. 

Hence, we will percolate not at a constant
above the threshold, but at a multiple above the threshold: $q=q_c/\gamma$.
We consider this problem in detail in a separate work\cite{sbr-percolation}.
The result is that if we {\em follow a random edge in the graph, the probability it
reaches an infinite component is $\delta=z/k_{max}$ for a constant $z$ which
depends only on $\tau$ and $\gamma$, but not $k_{max}$}.

Thus, since each high degree node has at least $k_{max}/2$ degree,
the average number of edges of a high degree node that
connect to the infinite component ($k_{inf}$) is at least:
\begin{equation}
k_{inf} \ge \delta \frac{k_{max}}{2} = \frac{z}{k_{max}}k_{max}/2
= \frac{z}{2} .
\nonumber
\end{equation}
The probability that a high degree node has at least one link to the
infinite component is at least:
\begin{eqnarray*}
P &\ge& 1 - (1-\delta)^{k_{max}/2}\\
&=& 1 - (1 - \frac{z}{k_{max}})^{k_{max}/2}\\
&\ge& 1 - e^{-z/2}.
\end{eqnarray*}
Thus both the average number of degrees that a high degree node
has to the giant component, and the probability that a high degree
node has at least one edge to the giant component are independent
of $k_{max}$.  So as we scale up $k_{max}$, we can expect that the
high degree nodes stay connected to the giant component.  We can
make $z$ larger by decreasing $\gamma$, particularly, if $1/\gamma
> 2/(3-\tau)$ we have $z>1$ \cite{sbr-percolation}.

It remains to be shown that the diameter of the connected
component is on the order of $O(\log N)$. To see this, we use the
approximate formula $l\approx \frac{\log M}{\log d}$ \cite{Newman}
of the diameter of a random graph with size $M$ and average degree
$d$.  We know that the size of the percolated graph is $\frac{N
z}{k_{max}}\langle k\rangle$ and that the average degree is
approximately 2\cite{sbr-percolation}.  Thus the diameter of the
giant component is:
\begin{eqnarray*}
l&=&\frac{\log( \frac{N z}{k_{max}}\langle k\rangle )}{\log(2)}\\
&=&\frac{\log \frac{N}{k_{max}} + \log z + \log \langle
k\rangle}{\log(2)}=O(\log N ) .
\end{eqnarray*}

At this point we have presented the main result.  If we can cache content on
high degree nodes, and query by percolation \emph{starting from} a high degree
node, we will always find the content we are looking for.  We have not yet
addressed how each node can find a high degree node.  In the next section we
show that by taking a short random walk through the network we will reach
a high degree node with high probability, and this gives us the final piece we
need to make the network searchable by all nodes.

\subsection{Random Walks Reach High Degree Nodes}
Consider a random PL network of size $N$ and with maximum node
degree $k_{max}$. We want to compute the probability that
following a randomly chosen link one arrives at a high degree
node. To find this probability, consider the generating function
$G_{1}(x)$\cite{sbr-percolation} of the degree of the nodes
arrived at by following a random link:

\begin{eqnarray}\label{1-3}
    G_{1}(x)=\frac{\sum_{k=2}^{k_{max}}k^{-\tau+1}x^{k-1}}{C},
\end{eqnarray}
where $C=\sum_{k=2}^{k_{max}}k^{-\tau+1}$. This results in the
probability of arriving at a node with degree greater than
$\frac{k_{max}}{2}$ to be:
\begin{equation}
\nonumber
    P_\tau=\frac{\sum_{k=k_{max}/2}^{k_{max}}k^{-\tau + 1}}{C} .
\end{equation}
Since the degrees of the nodes in the network are independent,
 each step of the random walk is an independent sample of the
same trial.  The probability of reaching
a high degree node within $\frac{\alpha}{P_\tau}$ steps
is:
\begin{eqnarray*}
1-(1-P_\tau)^{\alpha/P_\tau}&\ge&1 - e^{-\alpha} .
\end{eqnarray*}
Therefore, after $O(1/P_\tau)$ steps, a high degree node will
be encountered in the random walk path with high (constant) probability.
Now we need to compute $P_\tau$ for $\tau=2$ and $2 < \tau < 3$.
Since for all decreasing, positive,
$f(k)$ we have $\sum_{k=a}^b f(k) > \int_a^{b+1}f(k)dk > \int_a^{b}f(k)dk$
and $\sum_{k=a}^b f(k) < \int_{a-1}^b f(k)dk$,
we can bound the following sums.

If $\tau=2$, we have the probability of arriving at a node with
degree greater than $\frac{k_{max}}{2}$ is:
\begin{eqnarray*}\label{1-4}
    P_2&=&\frac{\sum_{k=k_{max}/2}^{k_{max}}k^{-1}}{C}\\
    &>& \frac{\log(k_{max})-\log(k_{max}/2)}{C}=\frac{-\log 2}{C},
\end{eqnarray*}
and
$C=\sum_{k=2}^{k_{max}}k^{-1}<\log(k_{max})$ . We finally get:
\begin{equation}\label{1-5}
    P_2>\frac{-\log 2}{\log(k_{max})} .
\end{equation}
For $\tau=2$, then in $O(1/P_2)=O(\log k_{max})$ steps we have reached a high
degree node.

If $2<\tau<3$, we have the probability of arriving at a node with
degree greater than $\frac{k_{max}}{2}$ is:
\begin{eqnarray*}
    P_\tau&=&\frac{\sum_{k=k_{max}/2}^{k_{max}}k^{-\tau+1}}{C}\\
    &>& \frac{1}{\tau-2}(2^{\tau - 2} - 1)\frac{1}{C k_{max}^{\tau-2}} ,
\end{eqnarray*}
and
$C=\sum_{k=2}^{k_{max}}k^{-\tau + 1}< \frac{1}{\tau-2}(1-\frac{1}{k^{\tau-2}})$.
We finally get:
\begin{equation}
    P_\tau>\frac{2^{\tau-2}-1}{k_{max}^{\tau-2}-1} .
\end{equation}
For $2< \tau < 3$, then in $O(1/P_\tau)=O(k_{max}^{\tau -2})$ steps we have
reached a high degree node, which is polynomially large in $k_{max}$ rather
than logarithmically large, as in the case of $\tau=2$.

A sequential random walk requires
$O(k_{max}^{\tau-2})$ time steps to traverse $O(k_{max}^{\tau-2})$
edges, and hence, the query implantation time will dominate the search time, making the whole search time  scale faster than $O(\log N)$. Recall that the percolation search step will only require $O(\log N)$ time, irrespective of the value of $\tau$.
A simple parallel query implantation process can solve the problem.
To implement $k_{max}^{\tau-2}$ query
seeds for example, a random walker with time to live (TTL) of
$K=\log_2 k_{max}^{\tau-2}$ will initiate a walk from the node in
question and at each step of the walk it implants a query seed,
and also initiates a second random walker with time to live $K-1$.
This process will continue recursively until the time to live of
all walkers are exhausted. The number of links traversed by all
the walkers is easily seen to be:
\begin{eqnarray*}
\sum_{i=0}^{K-1} 2^i &=& 2^{K} - 1\\
                     &=& k_{max}^{\tau-2} - 1 .
\end{eqnarray*}
Figure \ref{paral-vs-seq-fig} gives simulation results to show
that the parallel walk is effective, and thus search time scales
as $O(\log N)$ for all $2\le \tau < 3$. In practice, for values of
$\tau$ close to two, the quality of search is fairly insensitive
to how the number of query implants are scaled.

\subsection{Communication Cost or Traffic Scaling}
Each time we want to cache a content, we send it on a random walk
across $L(N,\tau)=O(1/P_\tau)$ edges.
When we make a query, if we reach the giant component, each edge passes
it with probability $q$ (if we don't reach a giant component only a constant
number of edges pass the query).  Thus, the total communications traffic scales
as $q E = q_c \langle k\rangle N/\gamma$.  Since $q_c=\langle k\rangle/\langle
k^2\rangle$ we have
$\mathcal{C_\tau}=O(\frac{\langle k\rangle^2 N}{\langle k^2\rangle})$.
For all $2\le \tau < 3$, $\langle k^2\rangle = O(k_{max}^{3-\tau})$.
For $\tau = 2$, $\langle k\rangle = \log k_{max}$ which gives
\begin{equation}
\label{eq:com_2} \mathcal{C}_2=O\left(\frac{\log^2 k_{max}
N}{k_{max}}\right)
\end{equation}
For $2<\tau<3$, $\langle k\rangle$ is constant
which gives
\begin{equation}
\label{eq:com_3} \mathcal{C}_\tau=O\left(k_{max}^{\tau - 3}
N\right)
\end{equation}
In section \ref{ssec:high_degree_nodes}, we showed that the number of
high degree nodes $H=O(N/k_{max}^{\tau-1})$.  We also know that
$L(N,\tau)=\alpha/P_\tau$ and $P_2 = O( 1/\log k_{max})$
and $P_\tau = O(1/k_{max}^{\tau - 2})$.
Thus we can rewrite the
communication scaling in terms of the high degree nodes,
$\mathcal{C}_\tau = O( L(N,\tau)^2 H )$.
So we see that communication costs scales linearly in $H$, but as the square
of the length of the walk to the high degree nodes.
This meets with our intuition since the high degree nodes are the nodes that
store the cache and answer the queries.

In the next section we discuss explicit scaling of $k_{max}$ to get
communication cost scaling as a function of $N$.
Tables
\ref{tb1} and \ref{tb2} show the scaling of the cache and communication cost
in $N$.
We see that for all
$\tau < 3$, we have sublinear communication cost scaling in $N$.

\subsection{On Maximum Degree $k_{max}$}
\label{ssec:on_maximum}
There are two ways to generate a random PL network:\\ (i)
 Fix a $k_{max}$ and normalize the distribution, i.e.,
\begin{eqnarray}\label{1-1}
    p_k&=&A k^{-\tau},0<k\leq k_{max}\ ,\\
    \hbox{where }\, A^{-1}&=&\sum_{k=1}^{k_{max}}k^{-\tau}\ .
\end{eqnarray}

To construct the random PL graphs,  $N$ samples are then drawn
from this distribution. For several reasons, the choice
$k_{max}=O(N^{1/\tau})$ is recommended in the literature
\cite{Aiello00}, and in our scaling calculations (e.g., Table
\ref{tb1})
we follow this upper bound.  \\
(ii) No a priori bound on the maximum is placed, and $N$ samples are drawn from
the distribution $p_k=A k^{-\tau}$, where
$A^{-1}=\sum_{k=1}^{\infty}k^{-\tau}$. It is quite straightforward to show that
almost surely, $k_{max} =O( N^{\frac{1}{\tau-1}})$. Thus,  when $\tau=2$,
$k_{max} =cN$ ($1>c>0$) in this method of generating a random PL graphs.

A potential problem with using the larger values of $k_{max}$, as
given by method (ii), is that the assumption that the links are
chosen independently might be violated. Random graph assumptions
can be shown to still hold when the maximum degree of a power-law
random graph is $k_{max}=O(N^{1/\tau})$ \cite{Aiello00}. This
however does not necessarily mean, that the scaling calculations
presented in the previous section do not hold for $k_{max} =O(
N^{\frac{1}{\tau-1}})$. In fact, extensive large-scale simulations
(see Section \ref{simul-sec}) suggest that one can indeed get
close to poly-logarithmic scaling of traffic (i.e., $O(\log^2N)$),
as predicted by the scaling calculations in this section.

There are several practical reasons for bounding $k_{max}$, as
well. First, in most grown random graphs, $k_{max}$ scales as
$N^{1/\tau}$. While grown random graphs display inherent
correlations, we would like to compare our scaling predictions
with performance of the search algorithm when implemented on grown
graphs. Hence, the scaling laws that would be relevant for such
P2P systems correspond to the case of bounded $k_{max}$. Second,
since the high degree nodes end up handling the bulk of the query
traffic, it might be preferable to keep the maximum degree low.
For example, for $\tau=2$, the traffic generated is of the same
order as the maximum degree, when $k_{max}=c\sqrt{N}$, thus
providing a balance between the overall traffic and the traffic
handled by the high degree nodes individually.

\begin{table}[htb]

\begin{center}

{\tt

\begin{tabular}{|c||c|c|c|c|}\hline

& Cache Size (TTL) & Query Cost\\\hline $\tau=2$ & O($\log N$) &O($\log^2
N$)\\\hline

$2<\tau<3$ & O($N^{\frac{\tau -2}{\tau -1}}$)
& O($N^{\frac{2\tau - 4}{\tau - 1}}$) \\\hline

\end{tabular}

} \caption{The scaling properties of the proposed algorithm when
$k_{max}=O(N^{\frac{1}{\tau-1}})$.} \label{tb1}

\end{center}
\end{table}

\begin{table}[htb]

\begin{center}

{\tt

\begin{tabular}{|c||c|c|}\hline

& Cache Size (TTL) & Query Cost \\\hline
$\tau=2$&O($\log N$)& O($\log^2(N) N^{1/2}$) \\\hline

$2<\tau<3$&O($N^{1-2/\tau}$)& O($N^{2-3/\tau}$)
\\\hline

\end{tabular}

}
\caption{The scaling properties of the proposed algorithm
when $k_{max}=O(N^{1/\tau})$.}

\label{tb2}

\end{center}
\end{table}

\section{Simulations on Random PL Networks}\label{simul-sec}
For all the simulations reported in this section, a random
power-law graph is generated with the method reported in
\cite{moll95}. The minimum degree of the nodes are enforced to be
two so that any node is part of the giant connected component with
probability one (see \cite{moll95}) . Note that in the
simulations, TTL refers to the length of the random walks
performed for content-list replication and query implantation. The
scaling enforced (if any) on the maximum degree ($k_{max}$) is
also reported for each simulation.

\subsection{Hit-rate vs. Traffic}

\begin{figure}
\begin{center}
\includegraphics[width=3.0in,height=2.0in]{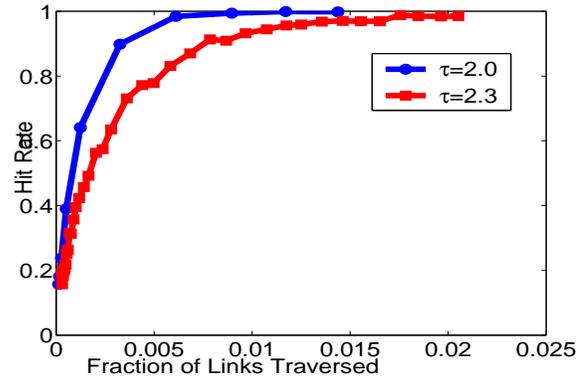}
\caption{The hit-rate as a function of the fraction of links used
in search, for networks with $\tau=2,2.3$. The number of nodes is
$30000$ and the TTL is 25 for both query and content implants. For
the case of $\tau=2$. For the case of $\tau=2$,  $k_{max}=2
N^{0.5} \approx 350$, while for the PL network with $\tau=2.3$ the
maximum degree is $k_{max}=2 N^{1/2.3} \approx 176$.
}\label{hitr-vs-traff-fig}
\end{center}
\end{figure}

Fig. \ref{hitr-vs-traff-fig} shows the hit rates achieved assuming
that each node has a unique content. As expected, for the same
traffic (i.e., the number of links used in the bond percolation
stage of the algorithms) the hit rate for $\tau=2$ is greater than
that for $\tau=2.3$. Some of the statistics for hit rates and
corresponding traffic are listed in Table \ref{tbl3}.

\begin{table}
\begin{center}
\begin{tabular}{|c|c|c|c|c|}
  \hline
  Hit Rate & 50\% & 75\% & 90\% & 98\% \\
  \hline
  Unique Replicas & 1.3e-3 & 2.4e-3 & 3.2e-3 & 6.8e-3 \\
  \hline
  10 Replicas & N/A & N/A& 2.0e-4 & 4.7e-4\\
  \hline

\end{tabular}
\caption{The fraction of edges (i.e., the ratio of the traffic
generated by the percolation search and the traffic generated by
the straight-forward search where queries are relayed on every
edge) involved in a search for various hit-rates when (i) Each
node has a unique content, and (ii) 10 replicas of each content
are distributed randomly in the network. The results are for a
power-law network with $\tau=2$, $N=30K$, and TTL=25 for both
query and content implants (see Figs. \ref{hitr-vs-traff-fig}, and
\ref{rept-cont-fig}).} \label{tbl3}
\end{center}
\end{table}

\begin{figure}
\begin{center}
\includegraphics[width=2.8in,height=2.0in]{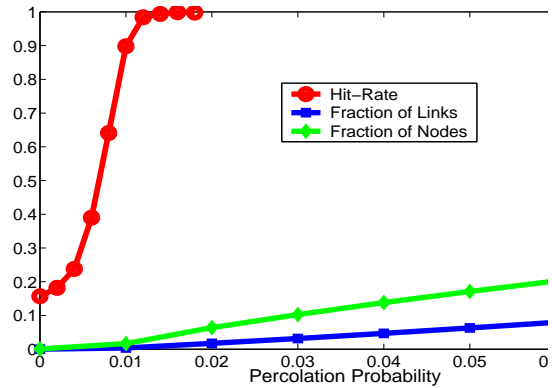}
\caption{The hit-rate, fraction of links and fraction of nodes
used in the search  as a function of the percolation probability
plotted together for comparison. While there is a sudden jump in
the hit-rate just above the percolation threshold (an indication
of a first order transition), the number of links and nodes
participating in the search increases much more gracefully (an
indication of a second order transition, also manifested in the
linear growth of these parameters just above the percolation
threshold). $\tau=2$ and number of nodes is $30000$ with
$k_{max}=400$.}\label{phase-trans-fig}
\end{center}
\end{figure}

Fig. \ref{phase-trans-fig} illustrates the {\em first-order phase
transition} of query hit-rates, as opposed to the second-order
phase transition of the size of the largest connected component,
as a function of the percolation probability. As noted in the
introduction, this first-order phase transition is a key aspect of
the proposed algorithm.

\begin{figure}
\begin{center}
\includegraphics[width=3.0in,height=2.0in]{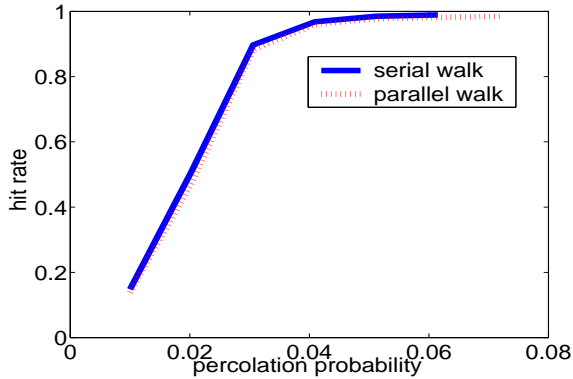}
\caption{Comparison of the hit-rate in parallel (circles ) and
serially (squares) implanted queries. In each case the total
number of queries are 16. The parallel implant uses four branching
random walks each of size $4$ and hence the total implantation
time is 8. While the serial implantation is a simple random walk
of size 16 and takes 16 time units. The network has $\tau=2.3$ and
$N=10000$. }\label{paral-vs-seq-fig}
\end{center}
\end{figure}

Fig. \ref{paral-vs-seq-fig} shows the performance of the search
algorithm, when the query-implantation step for the case of
$\tau>2$ is executed in parallel vs. when it is executed serially.
Recall that for $\tau>2$ the number of independent queries
required to ensure that one of the implanted queries is on a node
that is part of the giant connected component,  scales faster than
$O(\log N)$. Since, the query implantation time, if the
implantations were carried out by a serial random walk, would
dominate the desired search time of $O(\log N)$, we introduced a
parallel query implantation process (branching random walk), where
the walker constructs a binary tree, such that the total number of
nodes in the tree is the number of required query implantations.
As shown in Fig.\ref{paral-vs-seq-fig}, the performance of the
branching random walk is as good as a serial random walk.

\subsection{Repeated Trials}

\begin{figure}
\begin{center}
\includegraphics[width=3.0in,height=2.0in]{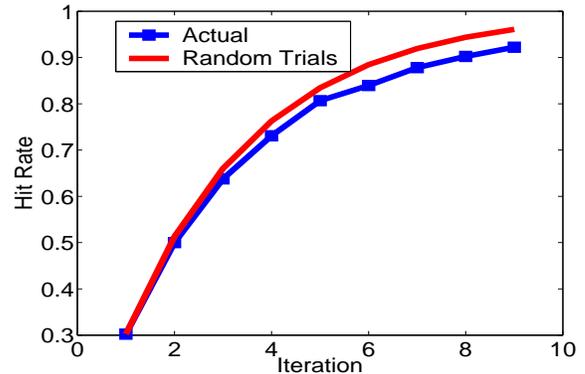}
\caption{The fraction of contents found as a function of the
number of times the search was repeated: Suppose a fraction $r$ of
contents were not found at the first try. If successive queries
were independent, the fraction of contents after the $K$'th try
should be around $1-r^{K}$. The actual fraction is plotted along
with what one expects from random tries. The network has size
$N=30,000$ and $\tau=2$. {\em TTL's are deliberately chosen to be
very low} (=5), so that $r$ is large
($>70\%$).}\label{repeats-fig}
\end{center}
\end{figure}

The results of Section \ref{III} guarantee that every content will
be found with probability approaching one, as long as the content
and query implantation steps are long enough. However, in
practice, at any percolation probability we will get a  hit rate
that is $< 1$, and the issue is what the behavior of the search
algorithm would be if one repeated the query a few times. If each
search is independent of the others, then we expect the hit rate
to behave as $1- (1-p)^r$, where $p$ is the hit rate for a single
attempt, and $r$ is the number of attempts. Fig. \ref{repeats-fig}
shows simulation results verifying this aspect of each query
attempt being almost independent of others. The fact that the hit
rate can be increased by repeated trials, is very important from
an implementation perspective: one does not need to know the
percolation threshold and the exact scaling of TTL's in order to
obtain very high hit rates. As shown in our simulations (Fig.
\ref{repeats-fig}), even {\em if we start with only a 30\% hit
rate}, the hit rate can be {\em increased to almost 90\% in only
seven attempts}.

\subsection{Content Replication}
\begin{figure}
\begin{center}
\includegraphics[width=3.0in,height=2.0in]{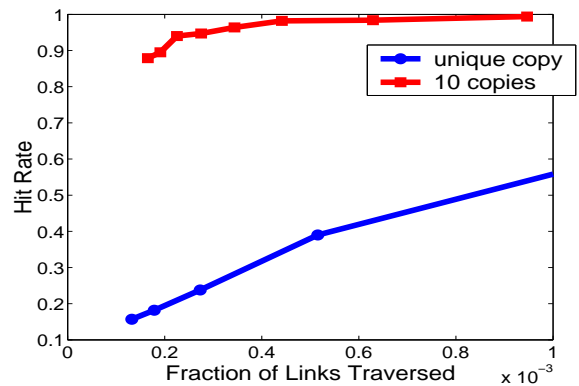}
\caption{The hit-rate as a function of the fraction of links used
in search, for $\tau=2$ for the case when 10 copies of each
content is in the network, along with the case of unique contents
for comparison. The number of nodes is $30000$, $k_{max}=375$ and
the average degree is $6$ and the TTL is 25 for both query and
content implants. }\label{rept-cont-fig}
\end{center}
\end{figure}

Next we consider another relevant issue: {\em what would be the
improvement in performance if multiple nodes in the network had
the same content}. As part of the percolation search algorithm, we
already execute a caching or a content implantation step that
makes sure that a subsequent query step would find {\em any
content} with probability approaching one. Now, if $l$ nodes share
the same content, then it will be implanted via $l$ different
independent random walks. In the case of random PL graphs, the $l$
different random walks for content implantation is equivalent to
looking for a content $l$ times independently (i.e., performing
the query implantation and bond percolation steps $l$ times),
while performing the content implantation random-walk only once.
Hence, in our percolation search algorithm, content replication
(due to nodes having the same content), {\em improves the hit rate
exponentially closer to 1}. The hit rates for unique vs. 10 copies
of contents are shown in Fig. \ref{rept-cont-fig}.

\subsection{Traffic Scaling}

\begin{figure}
\begin{center}
\vspace{.1in}
\includegraphics[width=3.0in,height=2.0in]{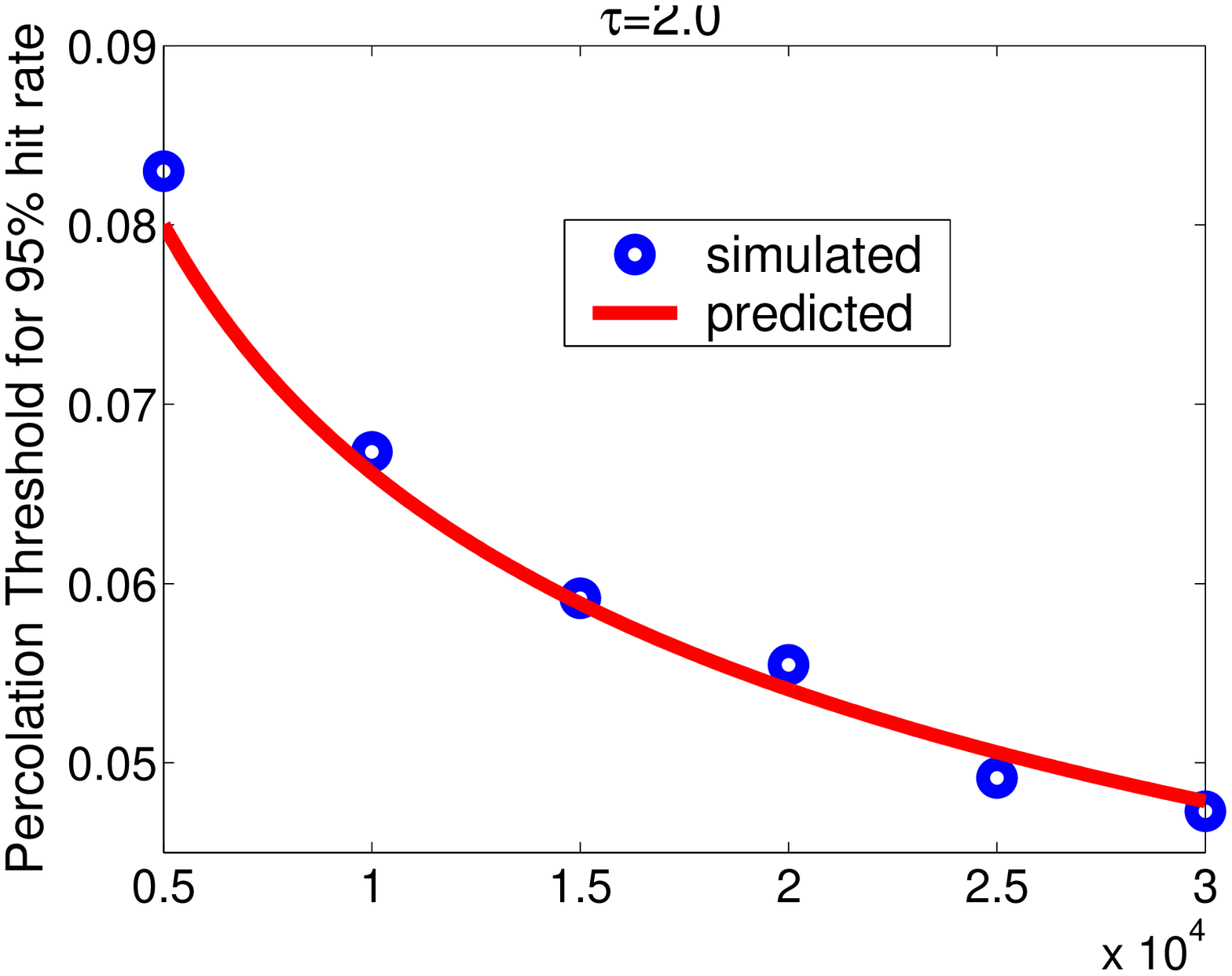}
\includegraphics[width=3.0in,height=2.0in]{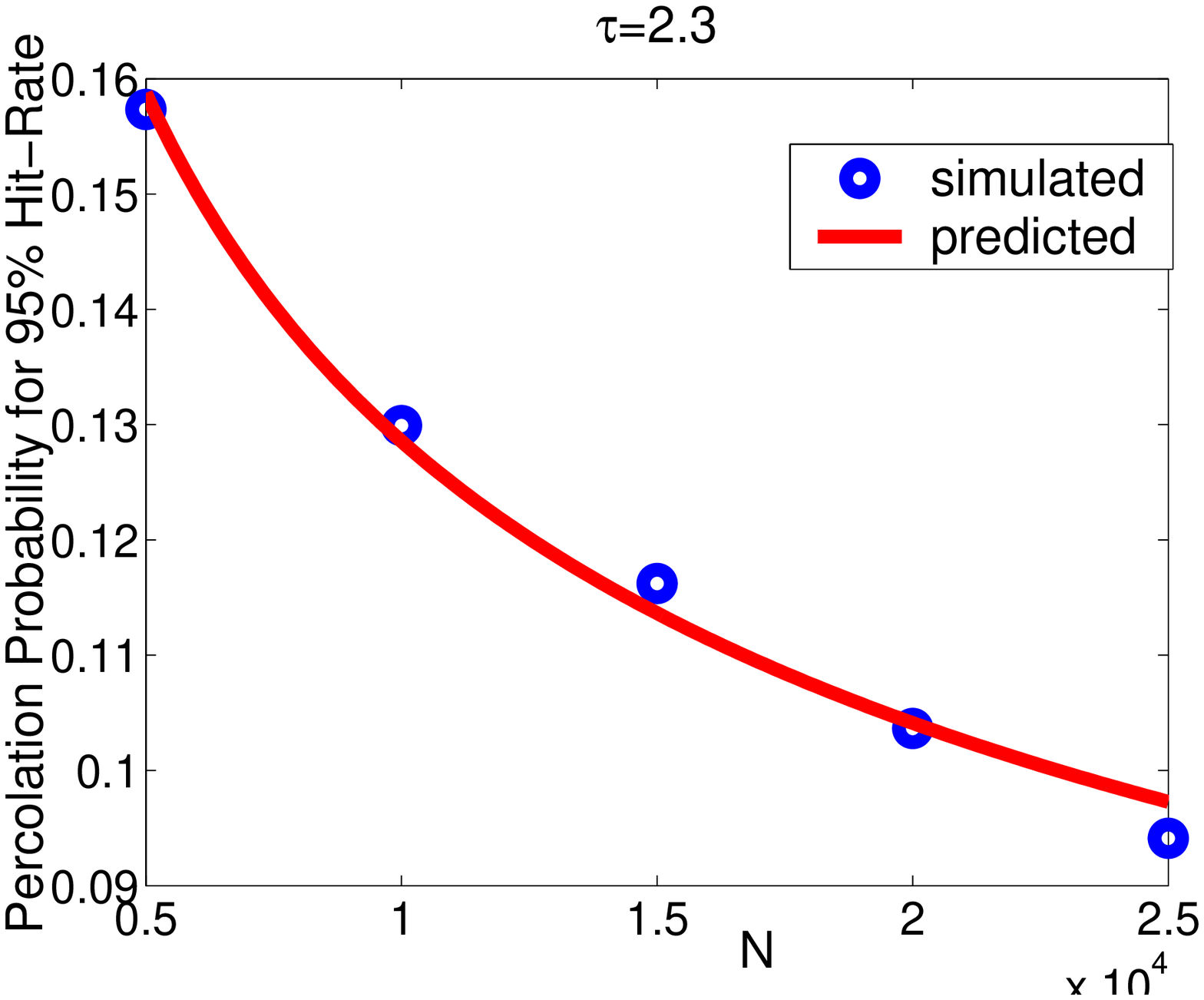}
\caption{Scaling behavior of the percolation probability required
for a fixed hit-rate of $95\%$ as a function of the network size
for networks with $\tau=2,\tau=2.3$. The TTL is increased
according to Table \ref{tb2} and the maximum degree is forced to
be $N^{1/\tau}$. The scaling predictions according to the Table
\ref{tb2} are also shown for comparison.}\label{scaling-fig-1}
\end{center}
\end{figure}

\begin{figure}
\begin{center}
\includegraphics[width=3.0in,height=2.0in]{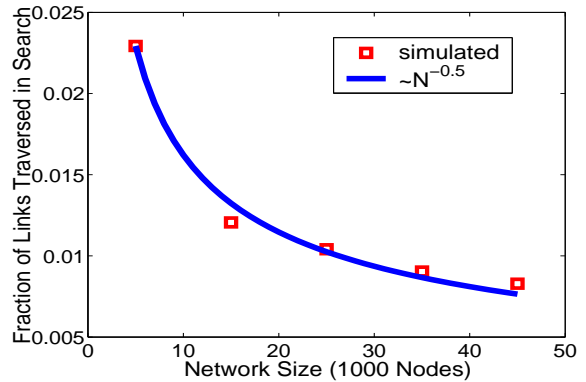}
\caption{Scaling behavior of the {\em fraction of links} required
for a fixed hit-rate of $90\%$ as a function of the network size
for a network with $\tau=2$. The TTL is increased according to
Table \ref{tb2} and the maximum degree is forced to be
$2\sqrt{N}$. The scaling is slightly improved to
$O\left(N^{-0.5}\right)$ for the fraction of links
required.}\label{scaling-fig-2}
\end{center}
\end{figure}

Fig. \ref{scaling-fig-1} shows actual scalings observed in our
simulations for various choices of $\tau$ and $k_{max}$. The
predicted scaling laws provide a good fit for the observed data
when $k_{max}$ is chosen to be $O(N^{1/\tau})$. The scaling for
the percolation probability required for a high hit rate matches
those predicted for the traffic reported in Table \ref{tb2}. On
the other hand, while Fig. \ref{scaling-fig-1} shows the scaling
of the percolation probability necessary to obtain a given target
hit-rate, the actual number of links traversed is in fact even
less. If each and every link had the chance to be traversed with
the percolation probability, then the actual traffic would
directly correspond to the percolation probability. A broadcast
started from a query implant, however, might end up at dead-end
nodes close to this implant. That results in {\em the actual
scaling of the traffic to be slightly better} than the scaling of
the required percolation probability. For $\tau=2$, for example,
the $O\left(\log^2 N/\sqrt{N}\right)$ scaling verified in Fig.
\ref{scaling-fig-1} has been modified to
$O\left(1/\sqrt{N}\right)$ as experimentally verified in Fig.
\ref{scaling-fig-2}.

\begin{figure}
\begin{center}
\includegraphics[width=3.0in,height=2.0in]{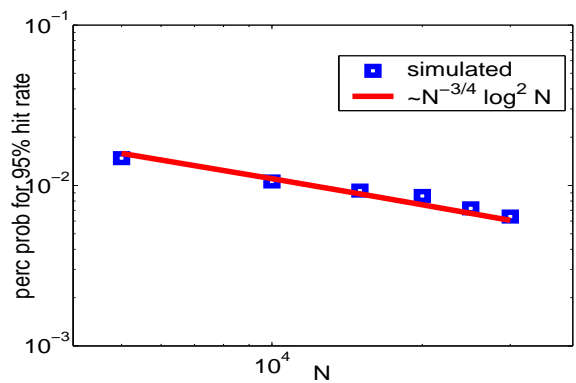}
\caption{Scaling behavior of the percolation probability required
for a fixed hit-rate of $95\%$ as a function of the network size
for a PL network with $\tau=2$, on a log-log basis. The TTL is
increased according to Table \ref{tb1}. The maximum degree however
is forced to be $2N^{3/4}$. The predicted scaling   is also
depicted for comparison.}\label{scaling-fig-3}
\end{center}
\end{figure}

\begin{figure}
\begin{center}
\includegraphics[width=3.0in,height=2.0in]{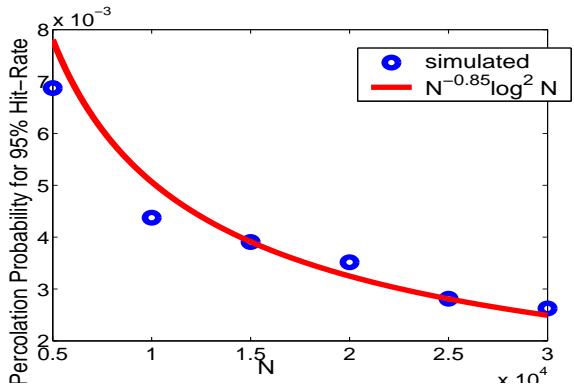}
\caption{The scaling of the percolation probability required for a
hit rate of $95\%$,when  $k_{max}=N/4$ and $\tau=2$ and $TTL=25$.
The scaling for $\log^2 N N^{-0.85}$ is also depicted for
comparison. It is important to note that simulations for such
large values of $k_{max}$ are fraught with difficulties. This
simulation however confirms the fact that while the scaling
results are precise when $k_{max}=O(N^{1/\tau})$ they still
closely match the simulations even in the extreme case of
$k_{max}=O(N)$.} \label{scaling-fig-4}
\end{center}
\end{figure}

More significantly, even when $k_{max}$ scales faster than
$N^{1/\tau}$, the same theoretical scaling laws seem to hold. As
an example of how the traffic scaling laws are we have provided
simulations for the case of $k_{max}\sim N^{3/4}$ (Fig.
 \ref{scaling-fig-3}), and $k_{max}\sim N$ (Fig.
 \ref{scaling-fig-4}) for $\tau=2$.

\section{Making Unstructured P2P Networks Scalable}\label{p2p-sec}
As noted in the introduction, a number of schemes have been
proposed to address the scaling problem in unstructured P2P networks, and the
following are a few of the more important ones:

\noindent
{\emph{ 1. Ultra-peer Structures and Cluster-Based Designs:}} A
non-uniform architecture with an explicit hierarchy seems to be
the quickest fix. This structure was also motivated by the fact
that the nodes in the network are not homogeneous;   a very large
fraction of the nodes have small capacity (e.g. dial-up modems)
and a small fraction with virtually unbounded capacity. The idea
is to assign a large number of low capacity nodes to one or more
Ultra-peers. The Ultra-peer knows the contents of its leaf nodes
and  sends them only the relevant queries. Among the Ultra-peers
they perform the usual broadcast search, where each query is
passed on every edge.

The Ultra-peer solution helps shield low bandwidth users; however,
the design is non-uniform, and an explicit hierarchy is imposed on
the nodes. In fact, the two-level hierarchy is not scalable in the
strict sense. After more growth of the network, the same problem
will start to appear among the Ultra-peers, and the protocol
should be augmented to accommodate a third level in the hierarchy,
and so on. In a more strict theoretical sense, the traffic still
scales linearly, but is always a constant factor (determined by
the average number of nodes per ultra-peer) less than the original
Gnutella system. Cluster-based designs \cite{cluster} are  more
centralized versions of practically the same idea, and therefore
suffer from the same issues.

Note that {\em the percolation search algorithm naturally distills
an Ultra-peer-like subnetwork} (i.e., the giant connected
component that remains after the bond percolation step), and no
external hierarchy needs to be imposed explicitly. Moreover, we
show in Section \ref{hetero-sec} that even if the random graph's
degree distribution is a mixture of two different distributions
(e.g., a heavy-tailed PL with $\tau\approx 2$, and a light tailed
PL with $\tau >4$), the percolation search algorithm naturally
shields the category of nodes with light-tailed degree
distribution, and most of the traffic is handled by the nodes with
heavy-tailed degree distributions.

\noindent {\emph{ 2. Random Walk Searches with Content
Replication:}} Lv et.al.\cite{Cohen} analyze random walk searches
with content replications, and their strategy is close to the work
of Adamic et. al, which was reviewed in Section \ref{prior-work} .
The idea is very simple: for each query, a random walker starts
from the initiator and asks the nodes on the way for the content
until it finds a match. If there are enough replicas of every
content on the network, each query would be successfully answered
after a few steps. In \cite{Cohen} it is assumed that a fraction
$\lambda_{i}$ of all nodes have the content $i$. They consider the
case where $\lambda_{i}$ might depend on the probability ($q_{i}$)
of requesting content $i$. They show that under their assumptions,
performance is optimal when $\lambda_{i}\propto \sqrt{q_{i}}$.

 This scheme has several
disadvantages. Since high connectivity nodes have more incoming
edges, random walks gravitate towards high connectivity nodes.
{\em A rare item on a low connectivity node will almost never be
found}. To mitigate these problems, \cite{Cohen} suggests avoiding
high degree nodes in the topology.

Moreover, this scheme is not scalable in a strict sense either:
even with the uniform caching assumption satisfied, the design
requires $O(N)$ replications per content, and thus, assuming that
each node has a unique content,  it will require a total of
$O(N^{2})$ replications and an average $O(N)$ cache size. The
above scaling differs only by a constant factor from the
straightforward scheme of all nodes caching all files. Finally, it
is a serial search algorithm, thus compromising the speed of query
resolution.

Clearly, the percolation search algorithm has several advantages
over this scheme and they are almost identical to the one's stated
in Section \ref{prior-work}, where the percolation search and the
random-walk based searches were compared. Moreover, the
percolation search algorithm finds any content, {\em even if only
one node in the network has it}, while the above algorithm relies
on the fact that a constant fraction of the nodes must have a
content, in order to make the search efficient.

\subsection{Percolation Search on Limewire Crawls}
We next address the issue of how well would the percolation search
algorithm work on the existing P2P networks. For our simulations
we have used a number of such snapshots taken by Limewire
\cite{limewire}. In particular, we have used snapshots number
1,3,5 from \cite{lime} with $N=64K,44K,30K$ respectively.

\begin{figure}
\begin{center}
\includegraphics[width=3.0in,height=2.0in]{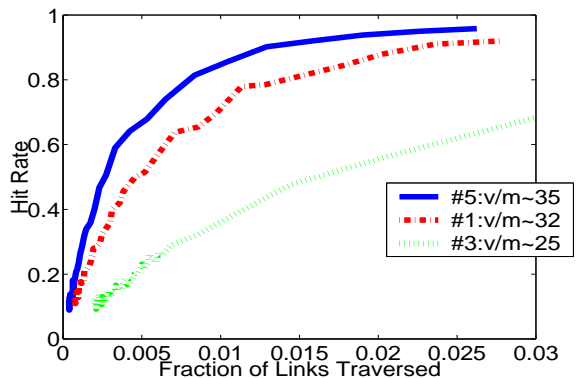}
\caption{The hit-rate as a function of the fraction of links used
in search, for Limewire crawl\#5, \#3, \#1. The ratio of the
variance to the mean for different crawls are indicated. The
performance of the percolation search algorithm is seen to be
dependent on this ratio, the higher the variance to mean, the
better the performance of the percolation search algorithm. The
TTL used for both query and content implant has length 25 for all
cases.}\label{lime-fig-1}
\end{center}
\end{figure}

\begin{figure}
\begin{center}
\includegraphics[width=3.0in,height=2.0in]{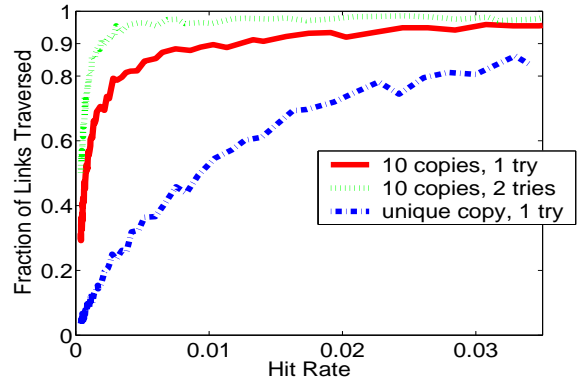}
\caption{For the Limewire crawl\# 5: the hit-rate as a function of
the fraction of the links required, when all contents are unique
as well as the case where 10 replicas of each content are in the
network. The case of two tries with 10 replicas is also quoted for
comparison.}\label{lime-fig-2}
\end{center}
\end{figure}

There are two important features about these snap-shot networks
that are relevant to
our discussions: \\
(i) Because of how one crawls the network, the resulting snap-shot
subnetworks are inherently networks obtained after bond
percolation, where the percolation probability is high but not
unity. The scaling laws of the percolation search algorithm
suggest that the performance of the
search on the actual graphs to be even better than those reported here.\\
(ii) The degree distributions of these networks are not ideal
power-laws, and at best they can be categorized as heavy-tailed
degree distributions.  A good measure of heavy-tailed degree
distribution is the ratio of the variance and the mean. In PL
networks with heavy tails, i.e., $2\le \tau \le 3$,
 this ratio is unbounded and goes to infinity as the network size grows.
  However, {\em one does not need these ideal conditions} for the percolation search algorithm to
  provide substantial reduction in traffic (see Figs. \ref{lime-fig-1} and \ref{lime-fig-2}).
  Recall that the search traffic generated in the percolation search algorithm is approximately
   $\langle k \rangle N q_c$, and hence is directly proportional to $\langle k \rangle q_c$.
We further know that $\displaystyle \langle k \rangle q_c \approx
\frac{\langle k \rangle^2}{\langle k^2 \rangle}$. Thus, as long as
the graph has a high root-mean-squared (RMS) to mean ratio, we
expect the percolation search algorithm to show substantial gains.
This is indeed the case in the implementations of our algorithm in
the crawl networks. Table \ref{tb4} shows that the overall traffic
can be reduced by 2 to 3 orders of magnitude without compromising
the performance.

\begin{table}
\begin{center}
\begin{tabular}{|c|c|c|c|c|}
  \hline
  Hit Rate & 50\% & 75\% & 90\% & 98\% \\
  \hline
  Unique Replicas & 3.1e-3 & 7.1e-3 & 1.3e-2 & 2.8e-2 \\
  \hline
  10 Replicas,2 tries & 1.1e-3 & 1.3e-3& 2.5e-3 & 6.3e-3\\
  \hline
  10 Replicas,1 tries & 1.3e-3 & 2.3e-3& 2.5e-2 & 4.6e-2\\
  \hline
\end{tabular}
\caption{For the Limewire crawl\# 5: the fraction of original
Gnutella traffic required for various hit-rates when all contents
are unique as well as the case where 10 replicas of each content
are in the network. The case of two tries with 10 replicas is also
quoted.}\label{tb4}
\end{center}
\end{table}

\section{Percolation Search in Heterogeneous PL Random Graphs}
\label{hetero-sec}
So far, we have assumed a
uni-modal heavy tailed distribution for the networks on which
percolation search is to be performed. In reality, however,
most networks are heterogeneous, consisting of categories
of nodes with similar capabilities or willingness to participate
in the search process; e.g., {\em the dominant categories
in existing P2P networks are, modems, DSL subscribers, and
those connected via high-speed T-1 connections}. Thus, the degree
distribution in a real network is expected to be a mixture of
heavy-tailed (for nodes with high capacity) and light-tailed (for
nodes with lower capacity) distributions. We now show that the superior
performance of the percolation search
algorithm is not limited to the case of a uni-modal power-law
random graph. In fact, as discussed before {\em the percolation search performs
well as long
as the variance of the degree distribution is much larger than its
mean}.

 Consider as an example, the case of a bi-modal network, where a
 fraction $x$ of the nodes have degree distribution $P_k$ with
 a heavy tail, while the rest have a light-tailed degree distribution
$Q_k$. Assume that the average degree of the two categories of nodes are
 the same for the sake of simplicity. The percolation threshold $q^{bi}_c$ of this graph is then related to
 $q_c$ the percolation threshold of a graph with the same degree
 distribution as of $P_k$ as: $q^{bi}_c \approx q_c/x$. Therefore, as long
 as a good fraction of all the nodes have a heavy tail, all
 observations of this paper still hold for a heterogeneous
 network. As far as the overall traffic is concerned, the total
 number of links traversed is at most $~(xN)p^{bi}_c=N p_c$ or the
 same as the case where all nodes had a heavy tailed distribution
 $P_k$. The query and content implantation times are however a bit
 longer in this case.

 Percolation search on heterogeneous networks, on the other hand, naturally
provides traffic shielding to low capabilities nodes. Consider
again a network with say two categories of nodes. The
 percolation search works by cutting out many links of the
 network, and {\em therefore almost all nodes participating in the search}
 process are the ones that are {\em highly connected}, which are almost surely
 part of the heavy tailed group. For instance, if the light tailed
 group has exponential degree distribution, then the probability
 of any of node of the light tailed category participating in the
 search process is exponentially small. Naturally then, {\em the nodes
 of the light tailed category are exempted from participation in
 the search process}.  See
the following table for a typical simulation result.
 \begin{table}[htb]
\begin{center}
\begin{tabular}{|c||c|c|c|}\hline
heavy tailed &light tailed &overall\\\hline\hline
3.50e-2&2.22e-5&6.12e-3\\\hline
\end{tabular}
{\caption{The fraction of \emph{nodes} that participated in a search
for a hit rate of $98\%$, in a network consisting of two power-law
modes: $4000$ nodes (called the heavy tailed mode) have a power-law
exponent $\tau=2$ while $20000$ others (called the light tailed
mode) have an exponent $\tau=4$. $TTL$ of $20$ was used for both
query and content implants.}

}
\end{center}\label{tb10}
\end{table}

\section{Concluding Remarks}\label{conc-sec}
We have presented a scalable search algorithm that uses
random-walks and bond percolation on random graphs with
heavy-tailed degree distributions to provide access to any content
on any node with probability one. {\em While the concepts involved
in the design of our search algorithm have deep theoretical
underpinnings, any implementation of it is very straightforward.}
Our extensive simulation results using both random PL networks and
Gnutella crawl networks show that unstructured P2P networks can
indeed be made scalable.

 Moreover, our studies show that even in networks
with different categories of nodes (i.e., graphs where the degree
distribution is a mixture of heavy-tailed and light-tailed
distributions) the search algorithm exhibits the favorable scaling
features, while shielding the nodes with light-tailed degree
distribution from the query-generated traffic. Our  recent
results \cite{hetero-protocol} indicate that it is indeed possible to have local rules,
that will enforce a desired category of the nodes in the network
to have either a heavy or light tailed degree distribution. One
can thus make sure that the subgraph consisting of the nodes with
low capacity has a light tail, and is thus exempted from the search
traffic with high probability. On the other hand, the high
capability nodes evolve into a subgraph with a heavy tail degree
distribution and hence will carry the majority of the search load.

Together with the new algorithms for building heavy-tailed growing
graphs, even in the presence of extreme unreliability of the nodes, and a
heterogeneous sets of nodes (in terms of connectivity and bandwidth capacities), the percolation search algorithm can provide an end-to-end solution for
constructing a large scale, highly scalable, and fault tolerant
distributed P2P networking system.


\end{document}